\documentclass[conference]{IEEEtran}

\usepackage{cite}
\usepackage{amsmath,amssymb,amsfonts}
\usepackage{graphicx}
\usepackage{booktabs}
\usepackage{textcomp}
\usepackage{xcolor}
\usepackage{url}

\begin{document}

\title{Sharing a Fabric with Collective Communication: Two Storage
Penalties in Deep Learning Training}

\author{\IEEEauthorblockN{1\textsuperscript{st} Chen Wang}
\IEEEauthorblockA{\textit{CCDS} \\
\textit{NTU Singapore}\\
Singapore \\
chen.wang@ntu.edu.sg}
\and
\IEEEauthorblockN{2\textsuperscript{nd} Wenzhao Wu}
\IEEEauthorblockA{\textit{CCDS} \\
\textit{NTU Singapore}\\
Singapore \\
wenzhao.wu@ntu.edu.sg}
\and 
\IEEEauthorblockN{3\textsuperscript{rd} Hyojin Kim}
\IEEEauthorblockA{\textit{CASC} \\
\textit{LLNL}\\
Livermore, USA\\
kim63@llnl.gov}
\and
\IEEEauthorblockN{4\textsuperscript{th} Jae-Seung Yeom}
\IEEEauthorblockA{\textit{CASC} \\
\textit{LLNL}\\
Livermore, USA\\
yeom2@llnl.gov}
}

\maketitle

\begin{abstract}
Distributed DL training on HPC systems often shares one network fabric
between NCCL/RCCL collective communication and parallel-filesystem I/O.
Using a real GNN training workload on a Slingshot-11 system, we show
that this sharing imposes \emph{two distinct costs}.
The primary cost is heavy-tailed DataLoader stalls: the typical DataLoader
wait is just 15\,ms at steady state, yet spikes to multiple seconds in
28\% of Lustre iterations and 12\% of VAST iterations.
The secondary cost is traffic-class contention on collective communication:
Lustre I/O stalls the all-reduce by up to 145$\times$ in an isolated
benchmark.
The two costs arise from different mechanisms. I/O stall latency affects any storage path that traverses the shared fabric, whereas all-reduce network contention occurs only when storage and collective communication share the same traffic class. Their common root cause is that storage I/O traverses the shared fabric. This work shows that node-local NVMe staging via DYAD\footnote{Our code is publicly available at \url{https://github.com/flux-framework/dyad}.} eliminates both effects by keeping storage I/O off that path. Across a full training epoch, DYAD achieves a 7.4$\times$ speedup over direct Lustre reads and a 1.06$\times$ speedup over VAST. By the second epoch, once the local cache is fully warmed, DataLoader stalls are eliminated entirely, allowing DYAD to reach a 1.31$\times$ speedup over VAST.
\end{abstract}

\section{Introduction}
\label{sec:motivation}
Modern HPC centers have shifted from a single parallel file system to tiered, protocol-specific storage because no single system can efficiently support the diverse I/O patterns of today's workloads~\cite{bez2023access,wang2021file}. Large-scale simulations require the sustained, high-bandwidth access provided by parallel file systems such as Lustre, whereas AI and data analytics rely on large numbers of small files accessed by heterogeneous clients. To support both, many centers deploy all-flash systems such as VAST alongside parallel file systems, often placing them on separate network fabrics. This architecture improves performance by matching workloads to appropriate storage tiers, but also makes data access performance more difficult to predict and optimize. In this work, we investigate the performance of a deep learning training workload in such a heterogeneous storage environment.

Multi-Instance Learning With Atomic Network (MILAN) \cite{Milan} is a structure-based deep learning framework that predicts protein–ligand binding affinity for drug discovery using equivariant graph attention networks. The model is trained on graph representations constructed from 3D protein–ligand complex coordinates and atomic features stored in HDF5 formats.
The training set used in this study, SAIR \cite{SAIR}, consists of 210 HDF5 files holding $\sim$10,000 complex structure entries each with up to 5 Boltz-1 \cite{Boltz} docking poses, for $\sim$5.2M total samples 
($\sim$157\,GB raw HDF5, plus a precomputed graph-edge cache of $\sim$529\,GB, $\sim$687\,GB total).
Training uses PyTorch Distributed Data Parallel (DDP) module across multiple
nodes and GPUs, with a \texttt{DistributedSampler}
that shuffles globally across all samples every epoch. In this study, we use four nodes and four AMD MI300A GPUs per node.

Each training iteration decomposes into four major stages, in order: a DataLoader wait for the next prefetched
batch (the slowest of the per-GPU I/O worker pool), a host-to-device (H2D) copy of the batched graphs onto the GPU, the forward pass including EGNN (Equivariant Graph Neural Networks), and the backward pass fused with the gradient all-reduce. 
Table~\ref{tab:steady-state} shows the cost breakdown at steady state using a all-flash VAST filesystem, with 4 nodes, 16 GPUs, and a batch size of 64. The median iteration takes $\sim$191\,ms, of which $\sim$68.6\% consists of the back-propagation (116\,ms) and the exposed DataLoader wait (15\,ms). The forward pass takes 18\,ms and the H2D copy takes 10\,ms. The rest is attributable to the optimizer and framework overhead. The DataLoader uses eight worker processes per GPU (PyTorch's standard
round-robin batching), each independently assembling one full 64-sample
batch, reading from two HDF5 files per sample: approximately 74\,KB of
raw atomic coordinates and features, and 212\,KB of graph edge data,
averaging 286\,KB per sample in total. Each worker's batch therefore
requires about 18\,MB of I/O; with the
throughput of a local flash tier, this completes well within the
144--214\,ms (median-p95)
GPU computation window, which is why the median exposed DataLoader wait is only 15\,ms at steady state.
\emph{The workload is therefore not bandwidth-limited but latency-sensitive: when the shared
network fabric introduces stochastic queuing delays, a single worker's fetch
can burst from a few milliseconds to several seconds, stalling the entire
batch.}

\begin{table}[!t]
  \centering
  \caption{The breakdown of steady-state per-iteration time (using 4 nodes,
  16 GPUs, and batch size 64). 
  }

  \label{tab:steady-state}
    \begin{tabular}{lrrr}
    \toprule
    Stage & Median & p95 & Max \\
    \midrule
    DataLoader wait & 15\,ms & 1{,}781\,ms & 7{,}134\,ms \\
    H2D copy & 10\,ms & 27\,ms & 70\,ms \\
    Forward pass & 18\,ms & 21\,ms & 359\,ms \\
    Backward + all-reduce & 116\,ms & 166\,ms & 3{,}315\,ms \\
    \bottomrule
  \end{tabular}
\end{table}

\begin{figure}[htbp]
  \centering
  \includegraphics[width=\linewidth]{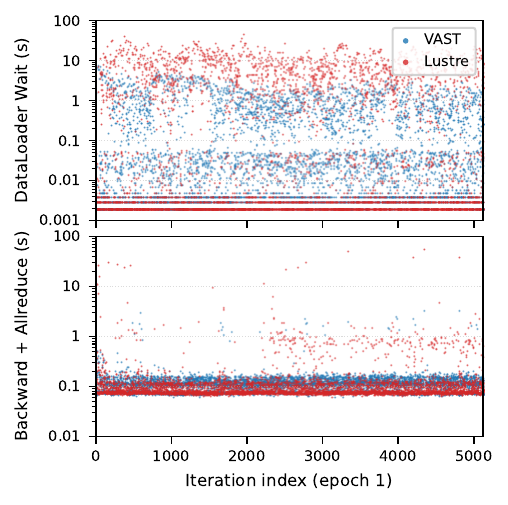}
  \caption{Per-iteration DataLoader wait time exposed (top) and backward pass +
  all-reduce time (bottom) from each of 5,120 iterations in a full
  training epoch. 
  }
  \label{fig:vast-lustre-intro}
\end{figure}

Figure~\ref{fig:vast-lustre-intro} shows the effect of storage tier on the
identical workload over a full epoch. Lustre takes 11,843.9\,s end-to-end
versus VAST's 1,693.9\,s, a 7$\times$ gap. Both panels show spikes, but the two spike types differ in character:
DataLoader stalls (top) are dense and sustained on Lustre while all-reduce communication spikes (bottom) are less frequent but still severe.
As Section~\ref{sec:interference} shows, these patterns arise from different
mechanisms, and understanding which one actually drives the wall-time penalty determines what kind of fix is worth building.

\section{Two Costs with Shared-Fabric Storage}
\label{sec:interference}

Routing storage traffic over the shared fabric introduces two distinct costs.
The \emph{primary cost} comes from contention with traffic external to the training job, resulting in heavy-tailed DataLoader latency that blocks the entire batch whenever a prefetch worker stalls. 
The \emph{secondary cost} arises from traffic-class contention among traffic originating from the same training job itself, which occasionally stalls the all-reduce. 
Unless otherwise noted, all measurements use the full MILAN training workload on Tuolumne, LLNL's production system.
We isolate the secondary effect first with a controlled microbenchmark, then
quantify both effects over a full training epoch.

\begin{figure}[htbp]
  \centering
  \includegraphics[width=0.6\linewidth]{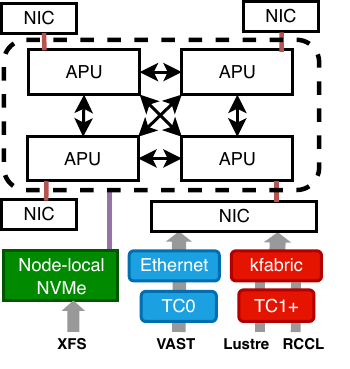}
  \caption{The system architecture of Tuolumne.}
  \label{fig:system-architecture}
\end{figure}

As shown in Figure~\ref{fig:system-architecture}, each compute node exposes four HPE Slingshot~11 Cassini NICs (CXI), each with two protocol faces on the same physical
hardware: an Ethernet face (TCP/IP) and a kfabric face (RDMA via the libfabric provider).
Lustre's connections and RCCL's Slingshot plugin both route over the
kfabric face and land in the same Slingshot traffic class (TC1+, HPC/RDMA).
VAST, mounted over NFSv3/TCP, uses the Ethernet face and lands in a separate
traffic class (TC0, Best Effort). The Slingshot fabric allocates injection
credits per traffic class (TC) per switch hop; TCs do not borrow credits from one another.

To investigate the secondary cost (the TC-contention mechanism on the all-reduce) in isolation, we built
a standalone benchmark\footnote{\url{https://github.com/psl-ntu/nccl_io_bench}}) that runs MILAN's
actual EGNN-based model and training-loop sequence (the real forward pass,
backward pass, and DDP gradient all-reduce) against each storage tier.
Each run performs 200 warmup iterations followed by 100 measured iterations
with no I/O (an uncontended baseline), then repeats the identical sequence
again with I/O enabled. 
We measure all-reduce costs over four storage tiers using the same 16-GPUs across 4 nodes, reporting the worst-case interference factor (with-I/O latency divided by without-I/O latency) in Table~\ref{tab:microbench}.

\begin{table}[htbp]
  \centering
  \caption{Worst-case all-reduce interference factor (with-I/O $\div$
  without-I/O) from the isolated
  microbenchmark.}
  \label{tab:microbench}
  \begin{tabular}{lr}
    \toprule
    Storage tier & Max factor \\
    \midrule
    shm (tmpfs, no network)         &   1.0$\times$ \\
    XFS (node-local NVMe)           &   1.0$\times$ \\
    VAST (NFS flash, separate TC)   &  24$\times$ \\
    Lustre (parallel FS, shared TC) & \textbf{145$\times$} \\
    \bottomrule
  \end{tabular}
\end{table}

In Table~\ref{tab:microbench},
\texttt{shm} and XFS show no detectable all-reduce interference. Their
storage paths do not touch the shared kfabric, so the all-reduce runs at baseline speed in every iteration. VAST's separate TC substantially
reduces interference, with a worst-case stall of 24$\times$. Lustre, sharing
TC with RCCL, reaches \textbf{145$\times$}, confirming that sharing a traffic
class with RCCL is the dominant driver of all-reduce degradation, consistent
with stochastic injection-credit contention.

The primary cost stems from per-sample storage
fetches over the shared fabric, which make the DataLoader latency distribution tail-heavy independently
of any RCCL contention. Table~\ref{tab:per-iter-stats} reports
per-iteration DataLoader wait and all-reduce statistics over a full
epoch with Lustre and VAST.
Median DataLoader wait is low on both (3\,ms on Lustre, 15\,ms on VAST), but both distributions are extremely
heavy-tailed. On Lustre, 28.2\% of iterations exceed 1\,s in
DataLoader wait (mean 2.0\,s, max 41.8\,s). VAST is not immune either: 11.5\% of iterations exceed 1\,s
(mean 324\,ms, max 7.1\,s), despite its storage traffic occupying a
separate traffic class. Traffic-class isolation moves the storage path
off the RCCL-shared TC but does not eliminate the DataLoader latency tail.

\begin{table}[htbp]
  \centering
  \caption{Per-iteration statistics over one full epoch.}
  \label{tab:per-iter-stats}
  \begin{tabular}{lrrrrrrr}
    \toprule
     & \multicolumn{3}{c}{DataLoader Wait} & \multicolumn{3}{c}{Backward + All-Reduce} \\
    \cmidrule(lr){2-4} \cmidrule(lr){5-7}
    Config & Mean & p95 & $>$1\,s & Mean & p95 & $>$1\,s \\
    \midrule
    VAST   & 324\,ms & 1{,}782\,ms & 11.5\% & 127\,ms & 166\,ms & 0.6\% \\
    Lustre & 1{,}987\,ms & 11{,}381\,ms & 28.2\% & 227\,ms & 474\,ms & 1.5\% \\
    \bottomrule
  \end{tabular}
\end{table}

To quantify how much each effect contributes to the epoch-time penalty,
we compute the per-iteration excess above the median for each of DataLoader wait and all-reduce latency,
then sum each across all iterations. For Lustre, DataLoader stalls
account for $\sim$93\% of the total excess iteration time versus only $\sim$7\% for
all-reduce spikes.
For VAST, DataLoader wait excess totals 1,608\,s versus only 119\,s for all-reduce excess.
Across every configuration we measured, DataLoader stall tail latency dominates the epoch-time penalty.



\subsection{Mitigations}
\label{sec:which-mitigation}

The two sources of costs require distinct solutions. 
All-reduce spikes reflect injection-credit
contention: they occur only when storage and collective traffic compete for
credits on the \emph{same} TC.
TC isolation (e.g., VAST on TC0) substantially reduces the secondary all-reduce effect,
but it does not fix the primary DataLoader stall effect. VAST's
DataLoader latency remains heavy-tailed because VAST's data path still traverses
the shared network fabric, just on a different TC that does not compete with RCCL.

Node-local NVMe hosting the entire training data addresses the root cause of both effects at once: storage
reads go through PCIe rather than any network fabric, so neither TC
contention nor fabric-credit exhaustion applies. As training data initially resides on
a shared parallel filesystem, utilizing node-local NVMe
requires a transparent staging layer to populate each node with its
required samples.
The I/O volume of this workload is modest: each sample is only $\sim$286\,KB, and a full 64-sample batch transfers about 18\,MB. The median DataLoader wait is only 3--15\,ms across configurations, indicating that sustained bandwidth is not the limiting factor. Instead, performance is dominated by heavy-tailed latency on the shared fabric. Staging the working set onto node-local NVMe removes these reads from the network path. At this I/O volume, local NVMe bandwidth is more than sufficient.
While a node-local NVMe is a promising solution, a single node-local NVMe often lacks sufficient capacity to host the entire training dataset. We therefore propose a solution that transparently manages sample locality across multiple node-local storage devices.

\section{DYAD}
\label{sec:granularity}

DYAD is a producer-consumer file-streaming system designed for HPC applications. A DYAD server service runs on each compute node, while the DYAD client library handles I/O requests through interception mechanisms (for C) or explicit API calls (for Python). Producers publish file metadata to a distributed KVS, and consumers query the KVS to identify the owning node before retrieving data via RDMA. In this workload, dataset shards are statically partitioned across nodes: each rank owns a disjoint subset of shards and serves byte-range requests from its assigned data.

However, DYAD and similar burst-buffer systems \cite{beegfs-caching,unifyfs} cannot be directly applied to the MILAN workload because they are designed around file-level staging. In DL training, the sample size is often orders of magnitude smaller than the containing file, and samples are not reused within the same epoch in general. Supporting this workload therefore requires adapting the staging granularity and timing. Our key design insight is to exploit the GPU computation phase between data consumption points 
to hide data movement, while ensuring that samples required for the next iteration are available before the subsequent forward pass begins.

\begin{figure}[htbp]
  \centering
  \includegraphics[width=\linewidth]{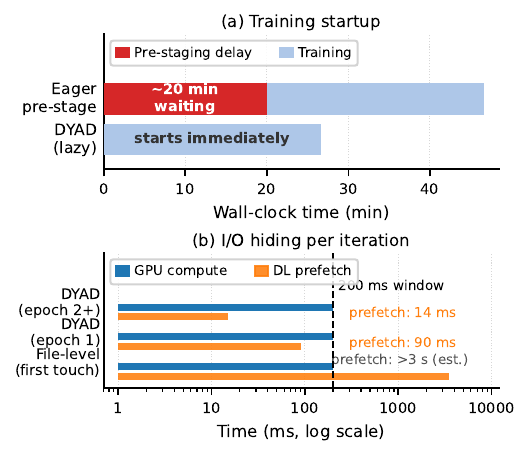}
  \caption{(a) Eager pre-staging imposes a $\sim$20-minute delay before
  training begins; DYAD starts immediately. (b) Per-iteration prefetch
  time (log scale) relative to the $\sim$144--214\,ms GPU compute window.
  File-level first-touch fetches far exceed the window, stalling the
  DataLoader; DYAD byte-range fetches stay hidden within it.}
  \label{fig:nimbus-design}
\end{figure}

Particularly, in the MILAN workload, each training iteration has a $\sim$144--214\,ms 
GPU compute phase during which the DataLoader prefetches the next batch. If every prefetch completes within this window, DataLoader wait approaches zero.
Achieving this requires two strategies working together, as explained below and illustrated in Figure~\ref{fig:nimbus-design}.

First, staging must be \emph{lazy}. DYAD fetches each sample's data
on demand during training rather than pre-staging all files before the run.
Eager pre-staging of the full $\sim$687\,GB dataset from Lustre would impose a
$\sim$20-minute delay before training could start, and would transfer bytes
that may never be accessed in a given run (e.g., a smoke test that didn't run for the full epoch).

Second, staging must be \emph{fine-grained}. Transferring whole files on
first touch would fetch several GB per shard, far exceeding the
$\sim$144--214\,ms compute window and stalling the DataLoader. By fetching
only the $\sim$18\,MB of blocks a worker's batch actually needs, the
first-touch latency stays within the window and the I/O cost is hidden.

\textbf{Byte-range extension.}
We extended DYAD with a byte-range fetch path: a new \texttt{dyad\_consume\_range()} client call
and a matching RPC handler allow a consumer to request an arbitrary
$[\mathit{offset},\,\mathit{offset}+\mathit{length})$ span, backed by a
lazy write-through cache of 64\,KB-block on node-local NVMe.
On a miss, the block-aligned span is read via \texttt{pread()} from the Lustre origin
and written to the local cache via \texttt{pwrite()}. A companion bitmap tracks resident
blocks so each block is fetched at most once per run. Concurrent requests
racing on the same block are safe under an exclusive lock (idempotent re-fetch is permitted). 
The lock covers only the bitmap update, not the fetch itself, to avoid
serializing on I/O latency.

\textbf{Dataset flattening.}
HDF5's per-sample access path resolves a group hierarchy and object header
on every access. Under global epoch shuffling, each \texttt{(pdbid, poseid)}
pair is read exactly once per epoch, so HDF5's metadata cache provides
no benefit. We apply a one-time preprocessing step that flattens the dataset
into a byte-indexed format: each sample's offset and length are precomputed
into an in-memory index, reducing every origin-side access to a single
\texttt{pread()} at a known offset with no per-sample metadata resolution.
This step completes in minutes and is required once per dataset.

\section{Evaluation}
\label{sec:dyad-eval}

We evaluate DYAD end-to-end against direct VAST and Lustre reads in the
native HDF5 format. All measurements
are a full single training epoch (5,120 iterations, $\sim$5.24M samples,
4-node/16-GPU/batch-64/8-workers-per-GPU), with DYAD using the MARGO~\cite{ross2020mochi} 
transport backend staging from a \emph{Lustre-backed} origin. All three
configurations process the identical global sample sequence each epoch.

\begin{table}[htbp]
  \centering
  \caption{Full single-epoch wall time.}
  \label{tab:full-epoch}
  \begin{tabular}{lrrr}
    \toprule
    Metric & VAST & Lustre & DYAD \\
    \midrule
    Full-epoch wall time (s) & 1{,}693.9 & 11{,}843.9 & \textbf{1{,}595.8} \\
    Throughput (samples/s) & 3{,}096 & 443 & \textbf{3{,}286} \\
    \bottomrule
  \end{tabular}
\end{table}
\label{sec:full-epoch}

Table~\ref{tab:full-epoch} shows that DYAD, staged from Lustre alone,
is 1.06$\times$ faster than direct VAST reads and 7.4$\times$ faster than
direct Lustre reads. In a first epoch, every sample is accessed exactly once,
so the NVMe cache provides no reuse benefit; the total data volume transferred
from Lustre is approximately equal to that of a direct Lustre read (64\,KB
block granularity may cause marginal overfetch, though optimizing block size
is left to future work). 
If a sample spans multiple 64\,KB blocks, the covering block-aligned span is fetched with a single synchronous \texttt{pread()}/\texttt{pwrite()}. Two factors contribute to the first-epoch advantage over direct Lustre.
First, DYAD uses a flattened data format rather than HDF5. Each sample access is reduced to a single \texttt{pread()} operation at a precomputed offset, eliminating the per-sample metadata traversal, group hierarchy lookup, and object-header resolution required by HDF5. 
Second, each file is assigned a single owner, ensuring that all sample fetches from a given file are issued by the same rank. This model reduces contention on the Lustre filesystem and improves the consistency of byte-range access performance. 
Together, flattened access and single-file ownership reduce per-sample I/O overhead and Lustre contention, enabling a worker's $\sim$18\,MB batch prefetch to complete within the GPU compute window. I/O is thus hidden behind computation, mitigating the DataLoader stalls that dominate the primary cost. 64\,KB alignment may add modest padding, but each miss remains a single contiguous \texttt{pread()}. 
This directly addresses the cause of the primary cost identified in Section~\ref{sec:interference}: only 2.93\% of DYAD iterations exceed 1\,s in DataLoader wait (150 of 5,120), versus 28.24\% for Lustre and 11.48\% for VAST. 
DYAD also mitigates the secondary cost. Lustre I/O and RCCL share the same Slingshot traffic class, so prolonged, many-client storage bursts contend with all-reduce for injection credits. In the first epoch DYAD still reads a comparable volume from Lustre, but single-owner flat \texttt{pread()}s replace many ranks' concurrent HDF5 metadata-heavy accesses to the same shards. Storage traffic on the fabric is therefore shorter-lived and less concurrent with collectives, reducing all-reduce interference. Empirically, the worst-case all-reduce time falls to 10.2\,s for DYAD in epoch~1 (versus 53.8\,s for Lustre and 3.3\,s for VAST. Together, these also explain why DYAD outperforms even direct
VAST reads: VAST substantially reduces all-reduce interference via TC isolation
but not the DataLoader stall tail, whereas DYAD eliminates both.


\subsection{Per-Iteration Interference}
\label{sec:per-iter}

\begin{figure}[t]
  \centering
  \includegraphics[width=\linewidth]{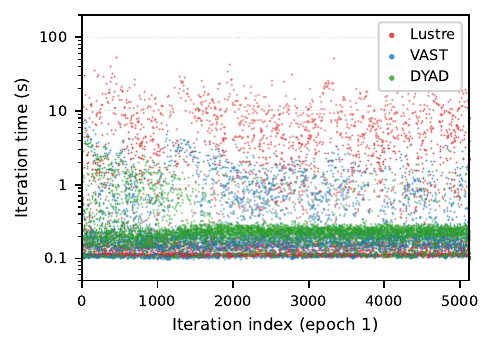}
  \caption{Total per-iteration time for every one of the 5,120 iterations
  in a full training epoch. Lustre's iterations are both more frequently and far more
  severely stalled, recurring throughout the entire epoch; DYAD, despite
  staging from the same Lustre origin, sits much closer to VAST.}
  \label{fig:scatter-bwd-nccl}
\end{figure}

Figure~\ref{fig:scatter-bwd-nccl} plots the total wall-clock time of
every logged iteration. VAST and DYAD both sit in a tight band around
200--400\,ms for the overwhelming majority of iterations; Lustre shares
a similar floor but has a visibly denser, heavier scatter of stalls
reaching into the tens of seconds throughout the entire epoch.
Specifically, 28.0\% of Lustre's iterations exceed 1\,s of total
iteration time (1,436 of 5,120), versus 9.9\% for VAST (505 of 5,120)
and 4.7\% for DYAD (243 of 5,120). Lustre stalls \emph{6$\times$ as
often as DYAD} and nearly $3\times$ as often as VAST. Lustre's worst iteration took 54.9\,s, versus 7.2\,s
for VAST and 10.3\,s for DYAD.

\begin{figure}[t]
  \centering
  \includegraphics[width=\linewidth]{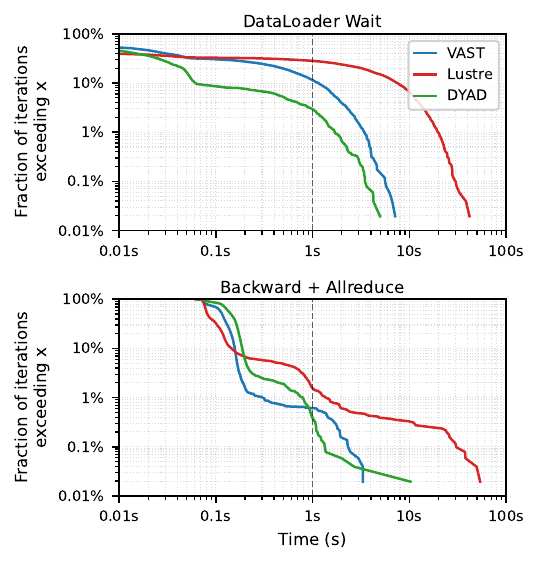}
  \caption{Tail distributions over the same full epoch, breaking total
  per-iteration time into DataLoader wait(top) and backward pass (bottom).
  Both are log-log ($y$ is the fraction of 5,120 iterations whose value
  exceeds $x$).}
  \label{fig:tail-bwd-nccl}
\end{figure}

\begin{figure*}[t]
  \centering
  \includegraphics[width=\textwidth]{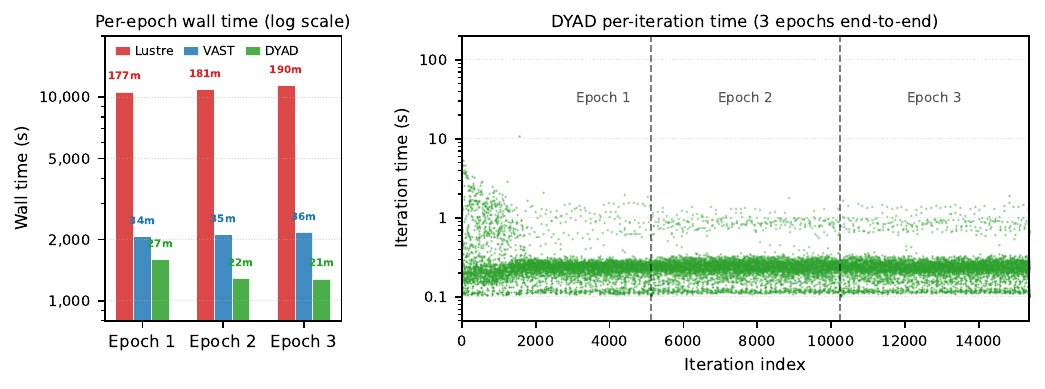}
  \caption{Left: per-epoch wall time for Lustre, VAST, and DYAD across
  three back-to-back epochs (log scale). VAST and Lustre show no
  inter-epoch improvement; DYAD drops sharply after epoch~1 as its
  node-local cache warms. Right: DYAD per-iteration time laid out
  end-to-end across all three epochs (dashed lines mark boundaries).
  The high-tail iterations in epoch~1 are first-touch Lustre fetches;
  they vanish in epochs~2 and~3 once the cache is fully warm.}
  \label{fig:epoch-comparison}
\end{figure*}

Figure~\ref{fig:tail-bwd-nccl} highlights the two most dominant impacts on the total iteration time during an epoch. The top panel (DataLoader wait) reveals the
primary cost: VAST carries a persistent shoulder in the 100\,ms--10\,s range
that DYAD does not, and Lustre's tail stretches an order of
magnitude further right than either. DYAD shows only 2.93\% of iterations
exceeding 1\,s in DataLoader wait, each plausibly a spike from a fresh 
fetch from the Lustre origin, a cost paid at most once per sample for the life of training. The bottom panel (backward + all-reduce) 
confirms the secondary effect: at the 1\,s line, Lustre sits at 1.54\%,
versus 0.61\% for VAST and 0.37\% for DYAD; the gap widens moving right. VAST
substantially reduces the secondary effect via TC separation, while DYAD
reducing storage traffic on the fabric reduces it further.
We note that DYAD is not immune to the fabric contention. Its
owner rank still \texttt{pread()}s each byte range's first touch from the
Lustre origin, over the same contended kfabric path RCCL uses. This is
visible in the top panel as DYAD's curve tracking slightly above VAST's
through the middle of the distribution before converging in the tail.
However, each span is fetched from Lustre \emph{at most once}, whereas direct Lustre reads pay the full fabric cost
on every sample access every iteration.

\subsection{Multi-Epoch Behavior}
\label{sec:second-epoch}

We ran DYAD for three consecutive full epochs (Figure~\ref{fig:epoch-comparison}).
By the second epoch, every sample DYAD needs is already resident on its owner's
node-local NVMe, so no further Lustre-origin fetches occur at all. 
Under global shuffling, a rank that needs a non-local sample retrieves it from the owner via RDMA rather than from the Lustre and
a third epoch confirms this is a stable plateau.
DataLoader wait drops from a first-epoch mean of 90.4\,ms (max 5.0\,s,
2.93\% of iterations exceeding 1\,s) to a second-epoch mean of 13.8\,ms
(max 901\,ms, \textbf{0\% of iterations exceeding 1\,s}), holding at a
third-epoch mean of 14.1\,ms (max 827\,ms, again 0\%). The DataLoader stall tail, the dominant contributor to epoch-time penalty, disappears entirely once the cache is warm.
All-reduce time stays roughly flat across all three epochs (mean
157.3\,ms $\to$ 171.4\,ms $\to$ 172.0\,ms), consistent with residual Lustre
traffic system-wide varying independently of DYAD's own cache state.

The first epoch already finishes in 1,595.8\,s, \emph{faster than direct
VAST reads' own single-epoch time} (1,693.9\,s, Table~\ref{tab:full-epoch}),
a 1.06$\times$ advantage, and epochs 2 and 3 grow this: 1,293.3\,s and
1,280.1\,s respectively, a 1.31$\times$ and 1.32$\times$ advantage over
direct VAST reads, and 7.4$\times$, 9.2$\times$, and 9.3$\times$ faster
than direct Lustre reads across the three epochs.

\section{Related Work and Conclusion}
\label{sec:conclusion}

\subsection{Related Work}

\textbf{Burst buffers and node-local aggregated storage.}
Burst buffer systems have been extensively studied~\cite{wang2016ephemeral,he2023hadafs,vef2020gekkofs,tatebe2022chfs}.
For instance, UnifyFS~\cite{unifyfs}
and BeeOND-based caching file systems~\cite{beegfs-caching} let a job
aggregate node-local storage into a temporary, POSIX-like shared namespace
for the duration of a run, primarily targeting checkpoint/restart and
shared-file HPC I/O patterns. DL-training-specific variants of this idea
include FanStore~\cite{fanstore2018}, which aggregates node-local storage
into a unified POSIX namespace via system-call interception; HVAC~\cite{hvac2022},
a distributed node-local read cache evaluated at 1,024-node scale; and
Monarch~\cite{monarch2022}, which transparently tiers data across node-local
storage and the parallel file system. However, all of these systems
typically operate at whole-file or POSIX-call
granularity and are not designed around a globally-shuffled,
sample-per-request access pattern. Moreover, none of them studies the interdependence
or relative magnitude of DataLoader-stall and collective-communication
interference effects, which this paper investigates.

\textbf{Caching and I/O systems for DL training.} Quiver~\cite{quiver}
introduces a cluster-wide, hash-addressed cache with substitutable cache
hits and job-aware prioritization, targeting cache efficiency across
concurrently running jobs sharing a dataset; DeepIO~\cite{deepio} uses
RDMA to shuffle and serve training samples from in-memory buffers spread
across node-local memory; CoorDL~\cite{coordl} coordinates data loading
and augmentation across the many jobs collocated on one server to cut
redundant preprocessing; DIESEL+~\cite{diesel2022} combines per-task
distributed caching with chunk-wise shuffling and GPU-assisted decoding
to accelerate image-dataset access; SHADE~\cite{shade2023} tracks
per-sample importance across a distributed job to improve the cache-hit
ratio under a fixed cache budget. Closer to our staging design,
DeepFetch~\cite{deepfetch2024} greedily prefetches samples from a
neighboring node's cache shard once local capacity is exhausted, and
NoPFS~\cite{nopfs2021} exploits the fact that the sampler's shuffling seed
makes an entire epoch's access sequence known in advance to schedule
near-optimal RAM/local-disk prefetching. These works optimize
data-loading throughput or cache efficiency, implicitly assuming a
storage path free of external interference. A
cache-efficiency-focused design does not by itself eliminate the
primary cost unless it also routes storage traffic off the shared fabric.

\textbf{HDF5-aware caching.} HDF5 Cache VOL~\cite{cachevol2022}
intercepts the HDF5 API to transparently cache data on node-local storage
and asynchronously migrate it to the parallel file system, hiding I/O
behind computation without application code changes. It targets the same
HDF5-on-node-local-storage setting as DYAD, but caches at the granularity
of HDF5 objects accessed through the library's own API. Under global
epoch shuffling, MILAN touches millions of distinct \texttt{(pdbid,
poseid)} groups exactly once per epoch, so caching at the VOL layer would
still pay HDF5's per-access group-hierarchy and object-header resolution
on every first touch. DYAD's dataset flattening instead resolves all
metadata once, offline, into a byte-offset index, avoiding this cost
entirely.

\subsection{Conclusion}

Routing storage I/O over the shared HPC
fabric imposes two distinct costs on distributed DL training. The primary cost, heavy-tailed DataLoader stalls, accounts for
$\sim$93\% of Lustre's epoch-time that is 7$\times$ larger than that of VAST. 
VAST also shows a similar DataLoader performance behavior (11.5\% of iterations stalling beyond 1\,s despite TC isolation).
The secondary cost, TC contention degrading the all-reduce by up to
145$\times$ in an isolated benchmark, accounts for $\sim$7\% of the
epoch excess and is substantially reduced by TC isolation. But fixing
only the secondary effect leaves the primary cost still high. Node-local NVMe, accessed via PCIe and thus invisible to the
fabric, eliminates both at once, if a staging system can get the right
bytes there transparently. DYAD does this via a lazy, block-granularity
byte-range cache over a flattened, byte-offset-indexed sample format,
staging only from Lustre and achieving 7.4$\times$
faster than Lustre and 1.06$\times$ faster than VAST in a first epoch,
growing to 1.3$\times$ in 2nd and 3rd epochs once the cache is warm and
first-touch fetches stop.


\section*{Acknowledgment}
This research is supported by the Ministry of Education, Singapore, under its Academic Research Fund Tier 1 (\#026632-00001, RS57/25), and by the NTU Singapore start up grant (\#025593-00001).
This work was performed under the auspices of the U.S. Department of Energy by Lawrence Livermore National Laboratory under Contract DE-AC52-07NA27344. LLNL-CONF-2022824

\bibliographystyle{IEEEtran}
\bibliography{references}

\end{document}